\documentclass{aa} 

\usepackage{graphicx}

\thesaurus{07 
          (07.13.1) 
          (02.16.1) 
          (02.19.2)}

\begin{document}

\title{On the interaction of radio waves with meteoric plasma}

\author{Luigi Foschini}


\institute{CNR -- Institute FISBAT, Via Gobetti 101, I-40129, Bologna 
(Italy); (email: L.Foschini@fisbat.bo.cnr.it)}

\date{Received 27 July 1998; accepted 2 November 1998}

\titlerunning{On the interaction of radio waves with meteoric plasma} 
\authorrunning{L. Foschini}

\maketitle

\begin{abstract}
In this paper, a meteoric plasma is analyzed from a physical 
viewpoint, with particular emphasis on its interaction with radio 
waves.  The attention is drawn to some macroscopic characteristics of 
a meteoric plasma and it is shown that the electron--ion collision 
frequency is not negligible, as commonly thought.  
\end{abstract}

\section{Introduction}
When meteoroids enter the Earth's atmosphere, they create long and 
narrow trails of ionized gas, which can scatter radio waves.  The 
meteor radio echo theory finds its roots in studies of the ionosphere 
made toward the end of the twenties (Skellet \cite{SKELLET}).  But only after the 
Second World War, and the development of military radar, the correlation between 
radio echoes and meteor trails became clear (Hey \& Stewart \cite{HEY}).  
The first experiments explicitly devoted to meteor studies were carried out 
by Pierce (\cite{PIERCE}), who observed Draconids during the night of 
9 to 10 October 1946.

The first theories on the interaction of radio waves with meteors were due 
to Lovell \& Clegg (\cite{LOVELL}), Kaiser \& Closs (\cite{KAISER}), 
Herlofson (\cite{HERLOFSON}).  Thereafter, between 1950 and 1960, a 
lot of efforts in this field were undertaken, but after 1960 the 
interest quickly decayed.  The state of knowledge was well exposed in a 
classical paper by Sugar (\cite{SUGAR}) and the book by McKinley 
(\cite{MCKINLEY}).  Toward the end of the eighties, the advent of 
digital technology renewed the interest in forward--scattering as a 
useful tool for communication channels over the horizon (see Weitzen 
\& Ralston \cite{WEITZEN}).  Today, meteor radars are widely used, 
even by amateur astronomers, because of low cost (Jenniskens et al.  
\cite{JENNISKENS}, Yrj\"{o}l\"{a} \& Jenniskens \cite{YRJOLA}).

Actually, the two main theories for radio echoes are due to Poulter 
\& Baggaley (\cite{POULTER}), who deal with back--scattering, and 
Jones \& Jones (\cite{JONES2}, \cite{JONES8}, \cite{JONES3}), who 
deal with forward--scattering.  We can note that both theories \emph{do not 
consider the meteor as a plasma}, but as a simply ionized gas, with a 
negligible collision frequency. This is an assumption quite common in 
work on radio echoes from meteors, except for Herlofson 
(\cite{HERLOFSON}), who first considered the meteor as a plasma. 

Concerning the difference between back--scattering and 
forward--scattering, it is worth noting that the two systems use 
different types of radio waves.  A forward--scattering radar uses a 
continuous sine wave, while a back--scattering one uses a pulsed wave.  
This influences the mathematical approach to the problem, but also 
physical theories.  The pulse shape of the back--scattering radar can be 
represented by a sum of several components with different frequencies.  
It follows that, during the propagation, the various components tend 
to change phase with respect to one another, which leads to a change 
in shape of the pulse.  The dispersion relation thus depends on the 
frequency and can be expressed in a Taylor series.  A detailed treatment 
of the pulse propagation in dielectrics is not the subject of this 
paper: the interested reader can find useful mathematical tools in, 
for example, Oughstun \& Sherman (\cite{PULSE}).  Here we want only 
to underline that we cannot simply consider the signal emitted by a 
back--scattering radar as sinusoidal: thus, the extension of the back--scattering 
echo theory to forward--scattering is not correct. 

The purpose of this paper is to settle some basic concepts in meteor 
physics and we leave a ``cook--book'' approach to other articles.  
The interaction of sine waves (for the sake of simplicity) in the radio 
frequency range with meteoric plasma is investigated. 
We will see that the common assumption about meteors as a collisionless 
ionized gas does not have any physical ground.

\section{Meteoric plasma}
As a meteoroid enters the Earth's atmosphere, it collides with air 
molecules.  At the heights where most meteors ablate, the mean free 
path of the air molecules is about $0.1-1$~m.  On the other hand, common 
meteoroid dimensions are of the order of $10^{-3}-10^{-2}$~m.  This 
means that there is no hydrodynamic flow around the meteoroid and 
single air molecules impact on the body.  If we consider a meteoroid 
a typical geocentric speed of 40~km/s, it can be found that air 
molecules impinge on the body with the same speed. The kinetic 
energy is about $1.3\cdot 10^{-18}$~J (8~eV) per nucleon: a nitrogen 
molecule then has an energy of about $3.7\cdot 10^{-17}$~J or 
230~eV.  The impact energy is readily transformed into heat, which makes 
atoms evaporate from the meteoroid.  The collisions between free atoms and 
air molecules produce heat, light and ionization, i.e.  a meteor.  
Since this transformation occurs throughout the flight, the meteoroid 
atoms are dispersed in a cylindrical channel along the path. The 
electron line density is proportional to initial mass of the 
meteoroid, because the air mass involved is negligible when compared 
to the meteoroid mass.

After the escape, the first collisions of meteoroid atoms with air 
molecules take place at a distance of about one mean free path 
from the meteoroid path. It is useful to consider only the first collision 
to be important for ionization. This explains why the radio echo 
quickly rises to maximum amplitude and then slowly decays.  At the 
moment of creation, all electrons are thus located inside a cylinder 
with a radius of about one mean free path.

It is possible to calculate the Debye length, a parameter that allows us 
to establish if the meteor is a plasma or simply an ionized gas.  If 
the Debye length is small when compared with meteor characteristic 
dimensions, then it is possible to speak of \emph{plasma}, i.e.  a gas 
where the electrostatic energy exceeds the thermal energy.  In this case, if the 
thermal energy produces deviations from charge neutrality, a 
strong electric field arises in order to restore the charge neutrality.  
On the other hand, if the meteor characteristic dimensions are small 
compared to the Debye length, this means that the thermal energy 
exceeds the electrostatic energy and there is no charge neutrality.  In 
this case we have a simple ionized gas and we cannot speak of a plasma.  
It is important to underline the difference between a plasma and an 
ionized gas: a plasma has some macroscopic properties, such as the 
Langmuir frequency, which are absent in ionized gas.

As known, the Debye length is obtained by equating the thermal energy to the 
electrostatic energy and equals (see Mitchner \& Kruger \cite{MITCHNER}):

\begin{equation}
	\lambda_{D}=\sqrt{\frac{\epsilon_{0}kT}{n_{e}e^{2}}} \ [{\rm m}]
	\label{e:debye}
\end{equation}

\noindent where $\epsilon_{0}$~[F$\cdot$ m$^{-1}$] is the vacuum 
dielectric constant, $k$~[J$\cdot$ K$^{-1}$] the Boltzmann constant, 
$e$~[C] the elementary electric charge, $T$~[K] the temperature and 
$n_{e}$~[m$^{-3}$] the electron volume density.  For the sake of 
simplicity, in this paper, the electron volume density is used, 
although in radar studies on meteors the electron line density is 
used.  However, it is easy to obtain the electron line density noting 
that the meteor trail is like a long circular cylinder.

Common radio meteors are characterized by electron volume densities 
between $10^{11}$~m$^{-3}$ and about $10^{20}$~m$^{-3}$ (Sugar 
\cite{SUGAR}) and temperatures between 1000 and 5000 K (Bronshten 
\cite{BRONSHTEN}, Borovi\v{c}ka \cite{BOROVICKA1}, Borovi\v{c}ka \& 
Zamorano \cite{BOROVICKA2}).  By substituting these values in 
Eq.~(\ref{e:debye}) it is possible to calculate the maximum value of the 
Debye length, which is about 0.01~m.  Comparing it to the lower value of the 
meteor initial radius (0.1~m), \emph{it is possible to say that the 
meteor is a plasma and not an ionized gas}.

Since any slight distortion of the plasma from a condition of 
electrical neutrality gives rise to strong restoring forces, we have to 
consider how fast these forces act.  Writing the equation of 
motion for the electrons, it is possible to find that they oscillate 
with a characteristic angular frequency (see Mitchner \& Kruger 
\cite{MITCHNER}):

\begin{equation}
	\omega_{p}=\sqrt{\frac{n_{e}e^{2}}{\epsilon_{0}m_{e}}} \ [{\rm 
	s}^{-1}]
	\label{e:pfreq}
\end{equation}

\noindent where $m_{e}$~[kg] is the electron mass.  It is possible to 
calculate Eq.~(\ref{e:pfreq}) for any type of charged particle, 
but as electrons move faster than ions, they give the main 
contribution to plasma frequency and other contributions can be 
neglected. The characteristic angular frequency is often called the 
Langmuir frequency.

Now, it is possible to understand the different behaviour of radio 
echoes.  Already Kaiser \& Closs (\cite{KAISER}) noted a different 
behaviour of meteors due to electron line density.  They introduced 
the names \emph{ underdense} and \emph{overdense} to characterize 
meteors with electron line densities, respectively, below or above the 
value $2.4\cdot 10^{14}$~m$^{-1}$.  In underdense meteors, the 
electron density is sufficiently weak to allow the incident wave to 
propagate along the ionized gas and the scattering is done 
independently by each individual electron.  On the other hand, when 
the meteor is overdense, the electron density is sufficient to reflect 
totally the incident wave.  It has been usual to resort to a simple 
model in which the trail is regarded as a totally reflecting cylinder.

\begin{figure}[t]
\centering
\includegraphics[scale=0.8]{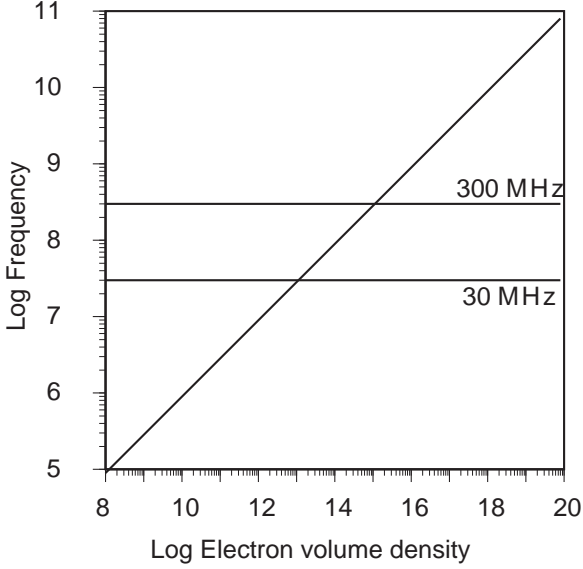} 
\caption{Plasma frequency as a function of 
electron volume density.  Typical radar frequencies are also shown 
for comparison. SI units are used.}
\label{FIG1}
\end{figure}

Now, comparing the electromagnetic wave frequency to the plasma 
frequency (Fig.~\ref{FIG1}) we can distinguish two regions depending on 
whether the plasma frequency is higher or lower than the radar 
frequency.  In the first case, the charges in the plasma have 
sufficient time to rearrange themselves so as to shield the interior 
of the plasma from the electromagnetic field (overdense meteor).  In the 
second case, since the radar frequency is higher the than plasma 
frequency, the incident wave can propagate along the plasma (underdense 
meteor). Moreover, space charges can appear.

Thus, a first method to distinguish underdense from overdense 
meteors, is by equating plasma frequency to radar frequency.  We stress
that this is valid for forward--scattering radar, which use continuous 
waves, while back--scattering radar see the meteors in a different 
way, because they use short pulses.

\section{General equations}
In a plasma the magnetic induction $\vec{B}$ and the magnetic field 
strenght $\vec{H}$ are almost always treated by the same 
relationship as in free space.  In order to account for polarizability 
associated with bound electrons of neutral particles and ions, it 
is necessary to calculate the dielectric constant of the plasma.  
However it is possible to see that, in the radar frequency range, the 
contribution due to polarizability is negligible.  Then, the relation 
between the electric field strenght $\vec{E}$ and electric induction 
$\vec{D}$ can also be treated as in free space.  The generalized Ohm 
law serves to relate $\vec{J}$ and $\vec{E}$:

\begin{equation}
	\vec{J}=\sigma\vec{E}=\frac{n_{e}e^{2}}{m_{e}(\nu 
	-i\omega)}\vec{E}
	\label{e:ohm}
\end{equation}

\noindent where $\nu$~[s$^{-1}$] is the mean collision frequency and 
$\sigma$~[S$\cdot$ m$^{-1}$] is the electric conductivity.
The meteoric plasma can be considered an isotropic medium since the 
electron gyrofrequency is much less than the radio wave frequency and thus 
the effect of the geomagnetic field can be neglected.

As stressed in Sect. 1, forward--scattering radars use continuous waves: 
taking into account that the time dependence of the 
electromagnetic field is $e^{-i\omega t}$, and using basic equations and 
standard identities, it is possible to combine Maxwell's 
equations in the following form:

\begin{equation}
	\nabla^{2}\vec{E}+(\omega^{2}\mu_{0}\epsilon_{0}+ 
	i\omega\mu_{0}\sigma)\vec{E} =\nabla(\nabla\cdot\vec{E})
	\label{e:e1}
\end{equation}

\begin{equation}
	\nabla^{2}\vec{H}+(\omega^{2}\mu_{0}\epsilon_{0}+ 
	i\omega\mu_{0}\sigma)\vec{H} =\nabla(\sigma 
	-i\omega\epsilon_{0})\times\vec{E}
	\label{e:h1}
\end{equation}

These are equations for the electric and magnetic fields in a meteoric plasma.  
It must be noted that Eqs.~(\ref{e:e1}) and (\ref{e:h1}) are useful in 
the underdense region, where there are space charges.  On the other 
hand, in the overdense region, when the plasma frequency is higher than the 
radar frequency, charge neutrality is present and then, the right side 
term of Eq.~(\ref{e:e1}) vanishes, according to Gauss law, when the net 
charge density is zero:

\begin{equation}
	\nabla^{2}\vec{E}+(\omega^{2}\mu_{0}\epsilon_{0}+ 
	i\omega\mu_{0}\sigma)\vec{E}=0
	\label{e:e2}
\end{equation}

The solutions of Eq.~(\ref{e:e2}) are nonuniform harmonic plane waves of 
this type:

\begin{equation}
	\vec{E}=\vec{E_{0}}e^{i(\vec{k}\cdot\vec{r}- \omega t)}
	\label{e:solu}
\end{equation}

\noindent where the wave vector in Eq.~(\ref{e:solu}) has the form:

\begin{equation}
	\vec{k}=\vec{\beta} +i\vec{\alpha}
	\label{e:wv}
\end{equation}

\noindent The two real vectors $\vec{\beta}$ and $\vec{\alpha}$ 
generally point in different directions.
If we substitute Eqs.~(\ref{e:solu}) and (\ref{e:wv}) in 
Eq.~(\ref{e:e2}), it is possible to find that:

\begin{equation}
	k^{2}=\omega^{2}\mu_{0}\epsilon_{0}+i\omega\mu_{0}\sigma
	\label{e:wave1}
\end{equation}

Taking into account Eqs.~(\ref{e:pfreq}) and (\ref{e:ohm}) and 
recalling that $\mu_{0}\epsilon_{0}=c^{-2}$ ($c$: light velocity in 
vacuum), it is possible to rearrange Eq.~(\ref{e:wave1}):

\begin{equation}
\begin{array}{l}
k^{2} = \frac{\omega^{2}}{c^{2}}\left\{1-\frac{(\frac{\omega_{p}}
{\omega})^{2}}{1+(\frac{\nu}{\omega})^{2}}+i\left[\frac{(\frac{\omega_{p}}
{\omega})^{2}(\frac{\nu}{\omega})}{1+(\frac{\nu}{\omega})^{2}}\right]\right\}
=\\[0.5cm]
       = \frac{\omega^{2}}{c^{2}}(\kappa_{R}+i \kappa_{I})
\end{array}
\label{e:wave2}
\end{equation}

The quantity $\kappa = \kappa_{R}+i \kappa_{I}$ is often identified as 
the \emph{complex dielectric constant} for the medium.
From Eqs.~(\ref{e:wv}) and (\ref{e:wave2}), when vectors $\vec{\beta}$ 
and $\vec{\alpha}$ point to the same direction, it is 
possible to define the real numbers $\alpha$ and $\beta$ as the {\it 
attenuation} and {\it phase} constants respectively:

\begin{equation}
	\alpha = \frac{\omega}{c}\sqrt{\frac{|\kappa|-\kappa_{R}}{2}}
	\label{e:alpha}
\end{equation}

\begin{equation}
	\beta = \frac{\omega}{c}\sqrt{\frac{|\kappa|+\kappa_{R}}{2}}
	\label{e:beta}
\end{equation}

If the collision frequency is negligible, i.e.  $\nu<<\omega$, then 
Eq.~(\ref{e:wave2}) can be reduced to:

\begin{equation}
	k^{2}=\frac{\omega^{2}}{c^{2}}\left\{1-(\frac{\omega_{p}} 
	{\omega})^{2}\right\}
	\label{e:wave3}
\end{equation}

\noindent that is, the equation used until now.  If 
$\omega>\omega_{p}$, then $k$ is a real number and the incident wave 
propagates without attenuation.  On the other hand, if 
$\omega<\omega_{p}$, then $k$ is a purely imaginary number and the 
incident wave is totally reflected.

\section{Collision frequency}
\begin{figure}[t]
\centering
\includegraphics[scale=0.9]{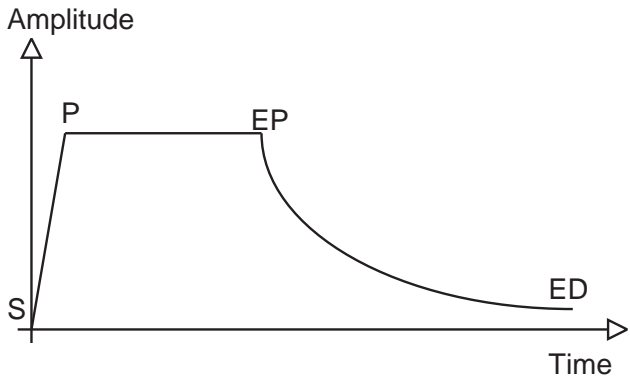} 
\caption{Typical radio echo from a meteor 
trail. The dimensions are exaggerated.  For an explanation of symbols, see 
text (Sect.~4).}
\label{FIG2}
\end{figure}

In all work on the theory of radio echoes from meteor trails, 
the collision frequency is considered 2 or 3 orders of magnitude lower than 
the radar frequency (a detailed analysis of past radio echo theories can 
be found in Foschini \cite{FOSCHINI2}).  This is the typical collision 
frequency of electrons with air molecules at the heights where meteors 
ablate.  Work on collisions in meteor trails generally deals 
with ionization and excitation, in order to know processes 
during trail formation (Massey \& Sida \cite{MASSEY}, Sida 
\cite{SIDA}, Baggaley \cite{BAGGALEY2}).  Other authors are interested 
in diffusion and thus study attachment, recombination and other 
chemical reactions between the atmosphere and meteoric plasma (Baggaley 
\cite{BAGGALEY1}, Baggaley \& Cummack \cite{BAGGALEY3}, Baggaley 
\cite{BAGGALEY2}, Jones \& Jones \cite{JONES9}).  In Fig.~\ref{FIG2} 
a typical overdense radio echo from a meteor trail is plotted.  When 
the ``flat top'' (P--EP) is reduced to zero, we have an underdense echo. 
Collision processes during the trail formation (from Start to Peak, 
S--P) are ionization and excitation, caused by air molecules impinging 
on the meteoroid. During echo decay, from End Peak (EP) to 
End Decay (ED), attachment and recombination are dominant processes, 
owing to the diffusion of the plasma in the surrounding atmosphere. 
During the \emph{plateau} (P--EP) the trail can be considered in 
\emph{local thermodynamic equilibrium}, when matter is in 
equilibrium with itself, but not with photons (Mitchner \& Kruger 
\cite{MITCHNER}). The plasma in the trail is transparent to radiation at 
some optical frequencies, which escapes from the meteor and form the 
light we observe.

For a plasma in local thermodynamic equilibrium it is still possible 
to employ the Boltzmann and Saha equations (see Mitchner \& Kruger 
\cite{MITCHNER}), even though complete thermodynamic equilibrium 
does not prevail.  Now, thermal ionization becomes the main process 
and metals, with low ionization energy, drive this phase.  It must be 
noted that in meteoroids there are some few per cent of alkaline and 
alkaline--earth metals, such as Na, K, Ca and Mg. 
These data are obtained from studies on meteorites (Mason 
\cite{MASON}, Millman \cite{MILLMAN}, Wasson \cite{WASSON}) and on 
bright fireball spectra (Bronshten \cite{BRONSHTEN}, Borovi\v{c}ka 
\cite{BOROVICKA1}, Borovi\v{c}ka \& Zamorano \cite{BOROVICKA2}).  
Because of their low ionization energy, at typical radar meteor 
temperatures (1000--5000 K), these metals easily ionize and thus, they 
are the main contributors to thermal ionization.  A small 
percentage of one of these metals suffices to produce a huge amount 
of electrons.

In the calculation of collision frequency of electrons with other particles, it 
must be taken into account that, because of the electrostatic field, ions have 
a cross section larger than that of atoms and molecules.  Thus, 
electron--ion collisions must be considered, and not electron--air 
molecule collisions, since the former are more frequent than any other.

Moreover, in calculating the electron--ion collision frequency we have considered 
potassium, because it has a low ionization energy (4.34 eV).  This 
choice could be questionable because potassium is scarcely present in 
meteor spectra (Bronshten \cite{BRONSHTEN}) or even absent 
(Borovi\v{c}ka \cite{BOROVICKA1}, Borovi\v{c}ka \& Zamorano 
\cite{BOROVICKA2}).  However, this metal is present in almost all 
meteoritic specimens (Mason \cite{MASON}) and other studies on meteor 
shower have reported its presence.  Specifically Goldberg \& Aikin 
(\cite{GOLDBERG}), using a rocket--borne ion mass spectrometer, 
detected the presence of potassium ion in $\beta$--Taurids.  The 
absence of potassium in meteor spectra is due to the low efficiency of 
light emission processes in this type of atom: it is well known that 
the most intense spectral line of potassium is in the infrared range 
(766.49~nm).  It is worth noting that also iron and sodium, two of 
the commonest elements in meteor spectra, could be absent in some 
cases (see Millman \cite{MILLMAN}).  
Therefore, we can consider a potassium percentage of about 1\% of 
meteoroid mass.

It is very difficult to find experimental values for 
electron--ion collision cross sections of various species, 
in the temperature range of meteors.  Experimental studies are mainly 
devoted to excitation and ionization cross sections, in order to 
compare results with data from spectra (Neff \cite{NEFF}, Boitnott \& 
Savage \cite{BOSA1}, \cite{BOSA2}, \cite{BOSA3}, Savage \& Boitnott 
\cite{SAVAGE}).
Rosa (\cite{ROSA}) gives collision cross sections of K, K$^{+}$ and 
some atmospheric gases, such as O$_{2}$, in the temperature range 
from 2000 K to 3500 K. It is important to note that 
the K$^{+}$ cross section is about 3 (\emph{three}) orders of magnitude 
larger than those of other species.

We have further supposed that only a single ionization is possible, 
i.e.  we have a reaction such as:

 \begin{equation}
	\mathrm{K}\leftrightarrow\mathrm{K}^{+}+\mathrm{e}^{-}
	\label{e:reaction}
 \end{equation}

We would like to stress that we are only interested in the analysis of
meteoric plasma in a steady state condition (during the \emph{plateau}) 
and we do not actually consider echo decay (recombination and 
other chemical processes) or trail formation (collisional ionization 
and excitation). Thus, it is possible to use a simplified model of 
collision frequency (Mitchner \& Kruger \cite{MITCHNER}), viz.:

\begin{equation}
	\nu_{ei}=n_{i}v_{ei}Q_{ei}
	\label{e:collf}
\end{equation}

\noindent where $Q_{ei}$ is the electron--ion collision cross section 
and $v_{ei}$ is the electron mean velocity with respect to ions (it is 
assumed to be about equal to the electron mean thermal velocity), and 
$n_{i}$ is the ion volume density.  In our case, owing to 
Eq.~(\ref{e:reaction}), $n_{i}=n_{e}$.

\begin{figure}[t]
\centering
\includegraphics[scale=0.8]{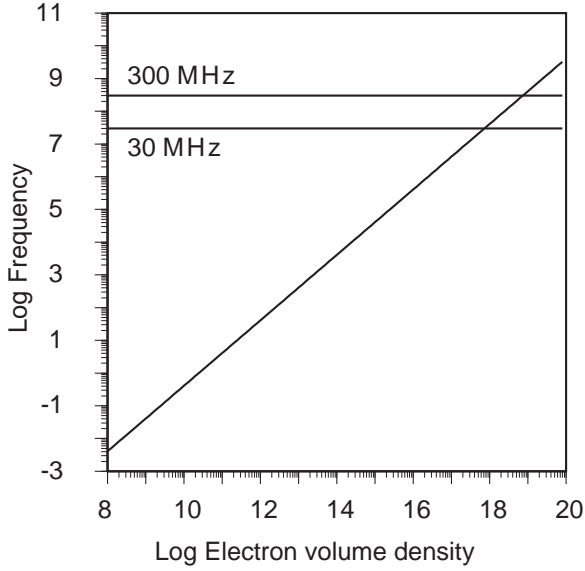} 
\caption{Electron--potassium ions collision 
frequency as a function of electron density.  Temperature 3000 K. Typical radar
frequencies are also shown for comparison.}
\label{FIG3}
\end{figure}

Now, it is possible to observe that $\nu_{ei}$ cannot be negligible 
anymore, especially for a high electron volume density 
(Fig.~\ref{FIG3}).  If we consider the contributions of other species, 
it is necessary to sum the various collision frequencies obtained 
using Eq.~(\ref{e:collf}).

Now, in the overdense regime, we have to consider the dispersion relation 
in Eq.~(\ref{e:wave2}) and not in Eq.~(\ref{e:wave3}). It is worth noting that 
Eq.~(\ref{e:wave3}) is still valid in the underdense regime because at high 
electron densities only the collisions are not negligible.

\section{Forward--scatter of radio waves from meteoric plasma}
It is possible to define, from Eq.~(\ref{e:wave2}), a \emph{critical 
electron density} in order to separate the underdense from the 
overdense regime and this occurs when:

\begin{equation}
	1-\frac{(\frac{\omega_{p}}{\omega})^{2}}{1+(\frac{\nu} 
	{\omega})^{2}}=0
	\label{e:critdens}
\end{equation}

\begin{figure}[t]
\centering
\includegraphics[scale=0.8]{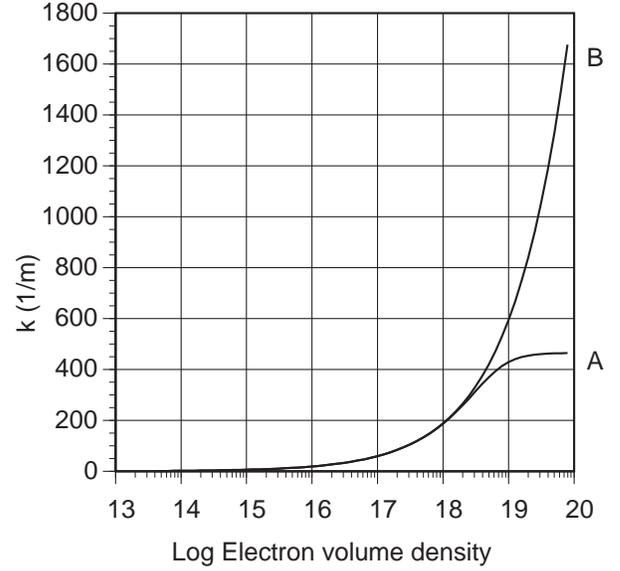} 
\caption{Wave vector modulus as a function 
of electron density above critical value.  Temperature 3000 K; Radar 
frequency 42.7 MHz. Line (A): with collisions; line (B): without collisions.}
\label{FIG4}
\end{figure}

\begin{figure}[t]
\centering
\includegraphics[scale=0.8]{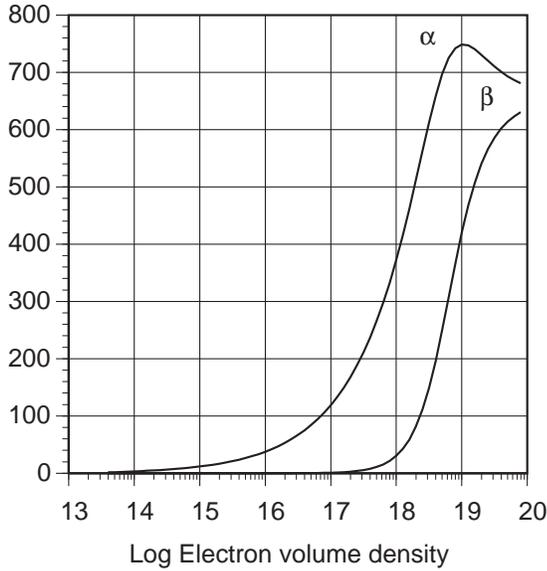} 
\caption{Attenuation and phase constants, 
taking into account collision frequency, as a function of electron 
density above critical value.  Temperature 3000 K; Radar frequency 
42.7 MHz.}
\label{FIG5}
\end{figure}

In order to make some example, we can assume a mean temperature of 3000~K. 
If we consider the CNR radar facility (Cevolani et al. \cite{CEVOLANI}), which
has a radar frequency of 42.7~MHz, we obtain a critical electron density of 
$n_{e}\cong 2.3\cdot 10^{13}$ m$^{-3}$. On the other hand, if we consider the 
M. de Meyere's radar (see C. Steyaert, Radio Meteor Obs. Bulletin, 
ftp://charlie.luc.ac.be/pub/icaros/rmob/), which has a frequency of 66.51~MHz, we
obtain a critical electron density of $n_{e}\cong 5.5 \cdot 10^{13}$ m$^{-3}$.
Below this critical density (underdense meteors), the wave number is real and 
the incident wave propagates into the meteoric plasma with negligible attenuation. 
We obtain well--known results about underdense meteors.  Above this value, 
waves which were previously excluded (see Eq.~(\ref{e:wave3})), can 
now propagate, but are strongly attenuated (Figs.~\ref{FIG4} and 
\ref{FIG5}).  From Fig.~\ref{FIG5}, it is possible to see that 
the presence of a collision frequency comparable with the radar frequency 
determines a rise of the real part of the wave vector modulus.  
Physically, this behaviour may be understood on the basis that 
collisions subtract energy from plasma oscillations, allowing the 
incident wave to penetrate, even if strongly attenuated.

We can observe two types and two sub--types of behaviour.  We have two 
main classes of meteors, \emph{overdense} and \emph{underdense}, 
according to whether the plasma frequency is higher or lower than 
the radar frequency respectively, depending on the initial electron volume 
density.  Moreover, overdense meteors are divided into two sub--classes 
(see Fig.~\ref{FIG5}): for an electron volume density between the 
critical density and about $10^{17}$ m$^{-3}$, the real 
part of the wave vector is negligible and the incident wave is 
totally reflected (\emph{overdense I}).  For a higher electron volume 
density, collisions allow the propagation of the incident wave, even if 
with a strong attenuation (\emph{overdense II}).

Total reflection, in the overdense I range, allow us to make some useful
approximations in calculating the attenuation of radio waves after the
forward--scattering. If the plasma had a definite 
boundary and a uniform electron density, then reflection at its 
surface would be simple because the gas would act as a dielectric with 
a complex dielectric constant.  But a meteor trail does not have a 
definite boundary and thus, the incident wave penetrates a little into 
the plasma before reaching the density necessary to allow total 
reflection. Reflection occurs gradually, as in a mirage.

\begin{figure}[t]
\centering
\includegraphics[scale=0.6]{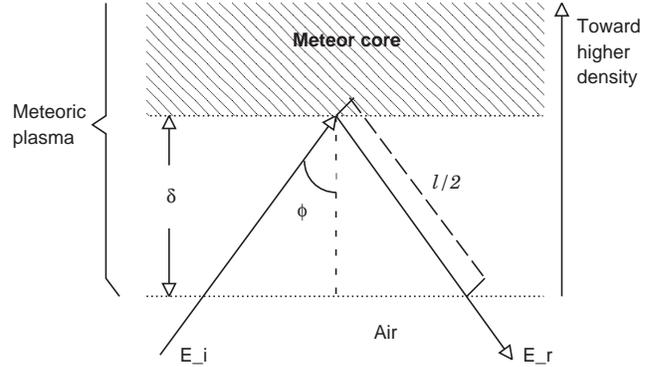}
\caption{Forward--scatter reflection geometry.}
\label{FIG6}
\end{figure}

Taking into account this fact, it is possible to make some 
approximations.  We can consider a simple geometry, as shown in 
Fig.~\ref{FIG6}, and then use the definition of the attenuation $a$ in 
decibel units:

\begin{equation}
	a=10\cdot \log \frac{|\vec{E_{i}}|^{2}} {|\vec{E_{r}}|^{2}} \ 
	[{\rm dB}]
	\label{e:decib}
\end{equation}

\noindent where subscripts $i$ and $r$ stand for incident and 
reflected wave. We substitute Eqs.~(\ref{e:solu}) and (\ref{e:wv}) in 
Eq.~(\ref{e:decib}) and, taking into account that the amplitude of a 
totally reflected wave is equal to the amplitude of the incident wave, 
we can obtain an attenuation value of about $a=-20\alpha l\log e$, where 
$l$ is the path of the wave into the plasma. From Fig.\ref{FIG6} we can see that:

 \begin{equation}
	l=\frac{2\delta}{\cos\phi}
	\label{e:path}
 \end{equation}

\noindent where $\delta$ is the penetration depth and $\phi$ is the 
incidence angle. Now, if we consider something similar to the ``skin 
effect'' in metals, we have $\delta = 1/\alpha$.  Then, 
Eq.~(\ref{e:decib}) becomes:

\begin{equation}
	a=\frac{-40\log e}{\cos\phi}\cong \frac{-17.36} {\cos\phi}=-17.36 
	\sec\phi \ [{\rm dB}]
	\label{e:decib1}
\end{equation}

The attenuation is simply a function of the angle of incidence and 
this is compatible with results obtained by Forsyth \& Vogan 
(\cite{FORSYTH}), who predicted an attenuation proportional to 
$\sec\phi$.  It is necessary to stress that Eq.~(\ref{e:decib1}) is valid only for 
type I overdense meteors, where total reflection is allowed.

In order to make some comparison with experimental data, we can 
consider a sample of radio echoes obtained from the 
CNR forward--scattering radar facility (e.g.  Foschini 
et al.  \cite{FOSCHINI}, Porub\v{c}an et al.  \cite{PORUBCAN}).  
Attenuation of reflected waves, in the overdense regime, ranges from 
-67 to -47~dB. By using the radar equation (Kingsley \& Quegan 
\cite{KINGSLEY}), attenuation values range from -155 to -125~dB, 
which gives a strong discrepancies with what is observed.
To obtain from Eq.~(\ref{e:decib1}) the attenuation values requested, it 
is necessary to use incidence angles from 68$\degr$ to 75$\degr$.  This 
is a very good approximation because the CNR radar has the 
main beam with about 15$\degr$ elevation angle: then 75$\degr$ is just
the complementary angle.

For overdense type I meteors, i.e. those with a \emph{plateau}, 
Eq.~(\ref{e:decib1}) can be used to calculate the meteor height.  
Taking into account the CNR radar geometry 
(transmitter--receiver distance is about 700 km, see Cevolani et al.
\cite{CEVOLANI}), it is possible to 
obtain a range of overdense type I meteor heights from about 94 to 
141~km.  This is very important, because the height is obtained 
without taking into account the diffusion coefficient, which is 
very uncertain. This requires further investigations, which are 
currently being carried out.

\section{Conclusions}
In this paper, we settled some basic concepts in meteor physics.  We 
have dealt with the interaction of sine waves, in the radio frequency 
range, with meteoric plasma. Attention is drawn to some 
macroscopic characteristics of a meteoric plasma and it is shown that the 
electron--ion collision frequency is not negligible, as commonly 
thought.  It is possible to define two meteor classes (overdense and 
underdense) according to whether the plasma frequency is higher or 
lower than the radar frequency respectively.  Overdense meteors are 
divided into two sub--classes (I and II), depending on the ratio between 
collision frequency and radar frequency.

Taking into account that the meteoric plasma does not have a definite boundary, 
a simple formula for the calculations of radio--wave attenuation after the 
forward--scattering is also presented.  This formula allows us to 
calculate the meteor height, without taking into account the diffusion 
coefficient.

Further questions can be put: in this work, potassium ion 
is considered, but other studies can be carried 
out in order to know the impact of other alkaline and alkaline--earth 
metals, such as Na, Ca and Mg.

\begin{acknowledgements}
Author wishes to thank G.~Cevolani, M.~de~Meyere and C.~Steyaert for technical
informations about their facilities.
\end{acknowledgements}


\begin{thebibliography}{}
\bibitem[1972]{BAGGALEY1} Baggaley W.J., 1972, MNRAS 159, 203

\bibitem[1980]{BAGGALEY2} Baggaley W.J., 1980, in: Proc.  IAU Symp.  
90: Solid particles in the solar system, ed.  I. Halliday and B.A. 
McIntosh, D. Reidel Pub.  Co., Dordrecht, p.  85

\bibitem[1974]{BAGGALEY3} Baggaley W.J., Cummack C.H., 1974, J. 
Atmos.  Terr.  Phys.  36, 1759

\bibitem[1970]{BOSA1} Boitnott C.A., Savage H.F., 1970, ApJ 161, 
351

\bibitem[1971]{BOSA2} Boitnott C.A., Savage H.F., 1971, ApJ 167, 
349

\bibitem[1972]{BOSA3} Boitnott C.A., Savage H.F., 1972, ApJ 174, 
201

\bibitem[1993]{BOROVICKA1} Borovi\v{c}ka J., 1993, A\&A 279, 627

\bibitem[1995]{BOROVICKA2} Borovi\v{c}ka J., Zamorano J., 1995, 
Earth, Moon, Planets 68, 217

\bibitem[1983]{BRONSHTEN} Bronshten V.A., 1983, Physics of meteoric 
phenomena, D. Reidel Pub.  Co., Dordrecht

\bibitem[1995]{CEVOLANI} Cevolani G., Bortolotti G., Franceschi C., Grassi G.,
Trivellone G., Hajduk A., Porub\v{c}an V., 1995, Planet.  Space 
Sci.  43, 765

\bibitem[1955]{FORSYTH} Forsyth P.A., Vogan E.L., 1955, Can.  J. 
Phys.  33, 176

\bibitem[1997]{FOSCHINI2} Foschini L., 1997, Propriet\`{a} di 
riflessione di onde elettromagnetiche dalla materia interplanetaria, 
Thesis in Physics, University of Bologna

\bibitem[1995]{FOSCHINI} Foschini L., Cevolani G., Trivellone G., 
1995, Nuovo Cimento C 18, 345

\bibitem[1973]{GOLDBERG} Goldberg R.A., Aikin A.C., 1973, Science 
180, 294

\bibitem[1946]{HEY} Hey J.S., Stewart G.S., 1946, Nature 158, 481

\bibitem[1951]{HERLOFSON} Herlofson N., 1951, Ark.  Fys.  3, 247

\bibitem[1957]{HINES} Hines C.O., Forsyth P.A., 1957, Can.  J. 
Phys.  35, 1033

\bibitem[1997]{JENNISKENS} Jenniskens P., Yrj\"{o}l\"{a} I., Sears P.,
Kuneth W., Rice T., 1997, WGN -- Journal of IMO 25, 141

\bibitem[1990a]{JONES2} Jones J., Jones W., 1990a, Planet.  Space 
Sci.  38, 925

\bibitem[1990b]{JONES8} Jones W., Jones J., 1990b, Planet.  Space 
Sci.  38, 55

\bibitem[1990c]{JONES9} Jones W., Jones J., 1990c, J. Atmos.  Terr.  
Phys.  52, 185

\bibitem[1991]{JONES3} Jones J., Jones W., 1991, Planet.  Space 
Sci.  39, 1289

\bibitem[1952]{KAISER} Kaiser T.R., Closs R.L., 1952, Philos.  Mag.  
43, 1

\bibitem[1992]{KINGSLEY} Kingsley S., Quegan S., 1992, 
Understanding radar systems, McGraw--Hill, London

\bibitem[1948]{LOVELL} Lovell A.C.B., Clegg J.A., 1948, Proc.  
Phys.  Soc.  60, 491

\bibitem[1971]{MASON} Mason B. (ed.), 1971, Handbook of elemental 
abundances in meteorites, Gordon and Breach Sci.  Pub., New York

\bibitem[1955]{MASSEY} Massey H.S.W., Sida D.W., 1955, Philos.  
Mag.  46, 190

\bibitem[1976]{MILLMAN} Millman P.M., 1976, in: Proc.  IAU Coll.  31: 
Interplanetary dust and zodiacal light, ed.  H. Els\"{a}sser and H. 
Fechtig, Springer--Verlag, Berlin, p., 359

\bibitem[1973]{MITCHNER} Mitchner M., Kruger C.H., 1973, Partially 
ionized gases, Wiley, New York

\bibitem[1961]{MCKINLEY} McKinley D.W.R., 1961, Meteor science and 
engineering, McGraw--Hill, New York

\bibitem[1964]{NEFF} Neff S.H., 1964, ApJ 140, 348

\bibitem[1997]{PULSE} Oughstun K.E., Sherman G.C., 1997, 
Electromagnetic pulse propagation in causal dielectrics, Springer, 
Berlin

\bibitem[1947]{PIERCE} Pierce J.A., 1947, Phys.  Rev.  71, 88

\bibitem[1995]{PORUBCAN} Porub\v{c}an V., Hajduk A., Cevolani G., Gabucci M.F.,
Foschini L., Trivellone G., 1995, Earth, Moon, Planets 68, 465

\bibitem[1977]{POULTER} Poulter E.M., Baggaley W.J., 1977, J. 
Atmos.  Terr.  Phys.  39, 757

\bibitem[1987]{ROSA} Rosa R.J., 1987, Magnetohydrodynamic energy 
conversion, Hemisphere Pub.  Co., Washington

\bibitem[1971]{SAVAGE} Savage H.F., Boitnott C.A., 1971, ApJ 167, 
341

\bibitem[1969]{SIDA} Sida D.W., 1969, MNRAS 143, 37

\bibitem[1931]{SKELLET} Skellet A.M., 1931, Phys.  Rev.  37, 1668

\bibitem[1964]{SUGAR} Sugar G.R., 1964, Proc.  IEEE 52, 116

\bibitem[1985]{WASSON} Wasson J.T., 1985, Meteorites -- Their record of 
early solar system history, W.H. Freeman and Co., New York

\bibitem[1988]{WEITZEN} Weitzen J.A., Ralston W.J., 1988, IEEE Trans. 
Ant. Prop. 36, 1813

\bibitem[1998]{YRJOLA} Yrj\"{o}l\"{a} I., Jenniskens P., 1998, A\&A 330, 739

\end{thebibliography}
\end{document}